# A Computational Pipeline for Patient-Specific Modeling of Thoracic Aortic Aneurysm: From Medical Image to Finite Element Analysis

Jiasong Chen[a], Linchen Qian[a], Ruonan Gong[a], Christina Sun[b], Tongran Qin[b], Thuy Pham[b], Caitlin Martin[b], Mohammad Zafar[c], John Elefteriades[c], Wei Sun[b], Liang Liang*[a]

[a] Department of Computer Science, University of Miami, Coral Gable, FL, USA

[b] Sutra Medical Inc, Lake Forest, CA, USA

[c] Aortic Institute at Yale-New Haven Hospital, Yale University School of Medicine, New Haven, CT, USA

## ABSTRACT

Thoracic aortic aneurysm (TAA) represents a critical cardiovascular challenge, characterized by the silent, progressive dilation of the aorta that can lead to catastrophic rupture or dissection. Despite the significant clinical burden of TAA, current clinical risk stratification relies heavily on maximum diameter measurements derived from three-dimensional computed tomography (3D CT), a metric that often fails to capture the complex biomechanical environment precipitating failure. To address this, we developed a comprehensive computational pipeline for patient-specific modeling of thoracic aortic aneurysms, combining segmentation, mesh generation, and finite element simulation. A key innovation of this pipeline is the open-arch mesh with node and element correspondence among different patients, enabled by the template-fitting based meshing method, which preserves anatomical accuracy, enables statistical shape modeling, and facilitates FE simulations with all-hexahedral elements in PyTorch-FEA —capabilities not addressed in previous studies. Our pipeline captures clinically relevant biomechanical differences, showing that aneurysmal patients exhibit significantly higher maximum absolute principal stress compared to non-aneurysmal patients. This result identifies a quantitative biomechanical marker that extends beyond conventional geometric measurements and potentially aids in TAA diagnosis. Moreover, the pipeline quantifies stress distribution across different regions of the aorta, enabling more detailed biomechanical assessment and supporting personalized treatment planning. By streamlining the transition from routine imaging to advanced stress analysis, this pipeline facilitates large-scale population studies and paves the way for the integration of patient-specific biomechanical parameters into routine clinical decision-making.

## 1. INTRODUCTION

The aorta is the body's largest arterial vessel, serving as the primary pathway for oxygenated blood within the systemic circulation[1]. Aortic aneurysms consistently rank among the top twenty causes of mortality in the United States[2]. Thoracic aortic aneurysm (TAA) arises from abnormal dilation of the thoracic aorta and remains a clinically significant disease, ranking as one of the leading causes of death in adults[3]. The estimated prevalence of TAA is approximately 1% in the general population[4,5]. TAA dissection happens when a tear forms in the inner layer of the aorta, resulting in the separation of the aortic wall layers[6]. A thoracic aortic aneurysm ruptures when the integrity of the aortic wall layers is compromised due to elevated blood pressure. If left untreated, both rupture and dissection are frequently fatal. Consequently, elective surgical repair is recommended for high-risk patients to prevent the aneurysm from rupturing or dissecting, involving replacing the aneurysm (i.e., part of the native aorta) with a vascular graft. While three-dimensional computed tomography (3D CT) remains the gold standard for diagnosing TAA, the evaluation of rupture and dissection risk is increasingly critical. This assessment relies on quantifying aortic geometry via medical imaging and analyzing wall stresses through finite element analysis (FEA).

The computational pipeline linking medical imaging to biomechanical analysis comprises three distinct stages: segmentation, mesh generation, and simulation. The initial stage, segmentation, generates volumetric masks that delineate the anatomical morphology of the target structure. Recently, deep learning–based image segmentation has emerged as a reliable method for extracting anatomical regions of interest from medical images[7–9]. For example, Cao[10] developed three convolutional neural network models based on the U-Net to simultaneously segment the entire aorta, true lumen (TL), and false lumen (FL), demonstrating the high accuracy and efficiency of U-Net based models for CT image segmentation of the aorta with type B aortic dissection. Cheung et al.[11] employs a 2D U-Net to segment the coronary arteries with high accuracy. Saitta[12] developed a fully automated pipeline that employs a 3D U-Net for thoracic aorta segmentation, subsequently utilize the segmentation results to perform geometric analysis of the aortic arch. However, raw segmentation outputs exhibit artifacts and surface irregularities that preclude their direct application in biomechanical simulations. Therefore, these voxel-based masks must be converted into structured mesh representations to ensure computational accuracy. Hexahedral meshes are commonly used in finite element simulations of the aorta due to their superior computational efficiency and accuracy[13]. Upon this discretized framework, the finite element method solves the governing partial differential equations (PDEs) to approximate the continuum mechanics of the aorta. This process allows for the faithful replication of real-world biomechanical behavior under specific physiological loading conditions. Due to anatomical variability, patient-specific modeling is essential for the detailed assessment of individual biomechanics behaviors, supporting precise simulations, accurate diagnoses, and personalized treatment strategies. Finite element (FE) simulations provide valuable insights into the biomechanical behaviors of tissues and organs in clinical studies. Developing accurate FE models represents a crucial initial step in establishing a patient-specific, biomechanically based framework for predicting the risk of TAA[14]. For example, Sun et al. observed distinct mechanical patterns in the aorta associated with dissection and subsequent repair through patient-specific finite element stress analysis[15].

The primary objective of this study is to establish a complete computational workflow for patient-specific modelling of thoracic aortic aneurysm that bridges the critical stages of image segmentation, mesh generation and finite element simulation.

## 2. METHODS

The proposed computational pipeline is illustrated in Figure 1, which consists of three main stages: segmentation, meshing, and simulation. The segmentation stage extracts the aorta geometry from 3D CT images, represented by a mesh with an arbitrary number of nodes and elements and a random topology. The meshing stage converts the segmented aorta geometry (i.e., a random mesh) into a quadrilateral surface mesh with a pre-defined topology, which is then converted to a solid hexahedral mesh with appropriate wall thickness. Finally, using the solid mesh, the simulation stage performs patient-specific FE simulations to obtain wall stress distributions. This pipeline connects raw CT images to the final patient-specific aortic wall stress assessment.

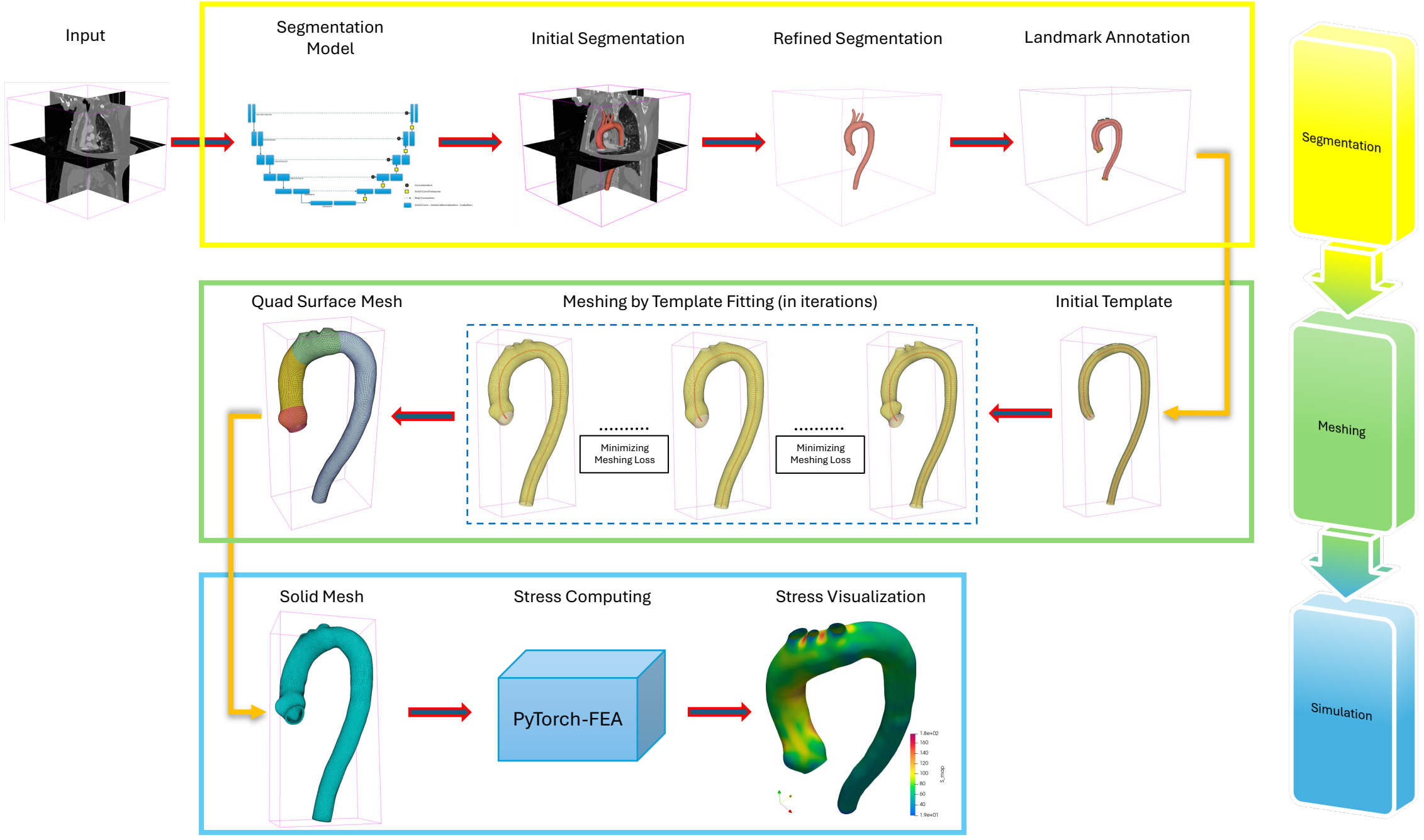

Figure 1. Pipeline Diagram

## 2.1 Dataset

This study utilized three fully de-identified datasets: one private and two public. The private dataset, acquired from Yale University – New Haven Hospital, includes 47 patients, comprises both individuals without TAA and those with TAA or Type A aortic dissections, with images acquired at multiple time points. To augment this cohort, we employed two public repositories: the HOCM dataset[16], contributing 27 CT scans from patients with a history of septal myectomy surgery, and the INSPECT dataset[17], a large-scale multimodal repository for pulmonary embolism. From INSPECT, a subset of 100 non-aneurysmal patients was randomly selected for this study.

## 2.2 Segmentation

The segmentation stage consists of three sequential steps: automated volumetric delineation via a deep learning model, manual refinement of segmentation masks, and expert annotation of anatomical landmarks.

Various deep learning-based segmentation models can be employed, such as nnU-Net[18], TransUNet[19], Swin UNETR[20], MedSAM[21]. This study utilized the nnU-Net[18] framework due to its state-of-the-art performance in medical imaging. The nnU-Net employs a U-shaped encoder-decoder configuration with skip connections, integrating a straightforward U-Net architecture within a highly optimized pipeline. The network comprises multiple resolution stages, where each computational block in the encoder and decoder follows a sequence of convolution, instance normalization, and Leaky ReLU activation. Training was optimized by minimizing a composite objective function consisting of Dice loss and standard cross-entropy loss, as defined in Eq. (1) and Eq. (2). Comprehensive architectural details are available in our previous work[22].

$$\mathcal{L} = \mathcal{L}_{Dice} + \mathcal{L}_{CE} \tag{1}$$

$$\mathcal{L}_{Dice} = 1 - \frac{2\sum_{i,j}\left(y(i,j)\hat{p}_m(i,j)\right)}{\sum_{i,j}\left(y(i,j)+\hat{p}_m(i,j)\right)} \tag{2}$$

Following automated segmentation, a rigorous manual refinement phase was conducted to correct artifacts and ensure high-fidelity aortic geometry. Three experts independently reviewed and iteratively adjusted the segmentation masks until consensus was reached. Upon finalization of the segmentation, anatomical landmarks—including hinge points, aortic arch curves, termination curve, centerline with maximal inner sphere radius—are manually annotated by three experts using the 3D slicer software[23]. The three hinge points define the hinge curve, serving as the inlet of the aortic geometry. The termination curve defines the outlet of the aortic geometry, located at the celiac trunk artery branch on a plane perpendicular to the descending aorta. Aortic arch curves are positioned approximately 2 mm above the contact points between the branch vessels and the aortic main body. The aortic centerline consists of 320 discrete points, and the maximal inner sphere radius was computed. These landmarks were essential for generating the initial quadrilateral mesh template in the subsequent stage. Figure 2 illustrates an example of the landmarks.

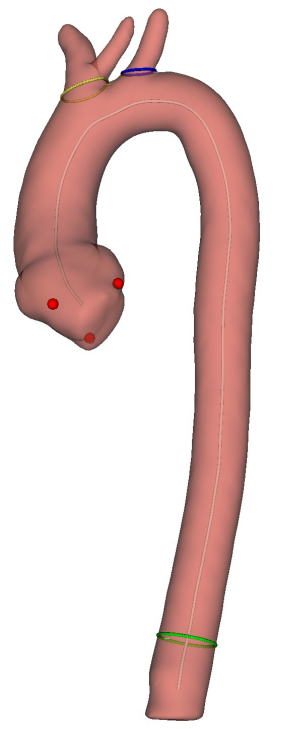

Figure 2. Landmark Annotation
Hinge points (red); Arch curves (yellow/blue); Termination curve (green); Centerline with maximal inner sphere radius (white)

## 2.3 Meshing

In the meshing stage, the process begins with the generation of an initial quadrilateral mesh template, which is subsequently deformed to match the aorta geometry obtained from the segmentation stage. This stage is fully automatic and implemented using the PyTorch library.

The initial quadrilateral mesh template is constructed as a tubular mesh. This tube is generated using 320 cross-sections perpendicular to the centerline, each centered on a point along the aorta centerline and with a radius equal to the corresponding maximal inner sphere radius. The centerline's start point is defined as the center of the three hinge points, and the end point corresponds to the end curve center. Each cross section (i.e., a circumference curve) consists of 78 evenly spaced points. The first cross-section lies on the hinge points plane and last cross-section lies on the end curve plane. To generate cross-sections ($G$), each centerline point ($C$) and its associated radius ($r$) establish a local coordinate system defined by the tangent vector ($T$), normal vector ($N$), and binormal vector ($B$). At each centerline point, each of the cross-section points ($P_i$) are generated according to Eq. (3). Subsequently, a quadrilateral mesh is constructed by connecting points between adjacent cross-sections. The arrangement of quad corners for vertices ($i$) in cross-sections $G_j$ and $G_{j+1}$ follows Eq. (4). Finally, the initial mesh template is completed by cutting holes in the aortic arch region of the mesh, corresponding to the patient-specific number of aortic arch branches.

$$P_i(\theta) = r_i cos\theta N_i + r_i cos\theta B_i + C_i , \theta \epsilon \{0,2\pi\} \tag{3}$$

$$Quad\ Corner\ (G_j, G_{j+1}) = (P_{i,j}, P_{i+1,j}, P_{i+1,j+1}, P_{i,j+1}) \tag{4}$$

In Eq. (4), $P_{i,j}$ denotes the point $i$ on the cross-section $j$. After generating the initial quadrilateral mesh template, a template fitting module deforms it to match the patient-specific aorta geometry obtained from the segmentation stage, using a designed meshing loss in Eq. (5) which measures the difference between the ground-truth mesh $S_{gt}$ and template mesh $S_{template}$, combining Chamfer loss with a mesh quality loss. The mesh quality loss includes element surface loss, surface flatness loss, element angle loss, and other element-quality related terms. Through iterative optimization, the template mesh deforms into a patient-specific quadrilateral aortic mesh, and the chamfer distance error between $S_{gt}$ and $S_{template}$ becomes sufficiently small.

$$meshing\ Loss = chamfer_loss(S_{gt}, S_{template}) + mesh_quality_loss(S_{gt}, S_{template}) \tag{5}$$

Figure 3 shows an example of deformed quadrilateral aorta mesh. Since the node correspondence, the deformed quadrilateral aorta mesh can automatically split into different aortic regions, including the aortic root, ascending aorta, aortic arch, and descending aorta.

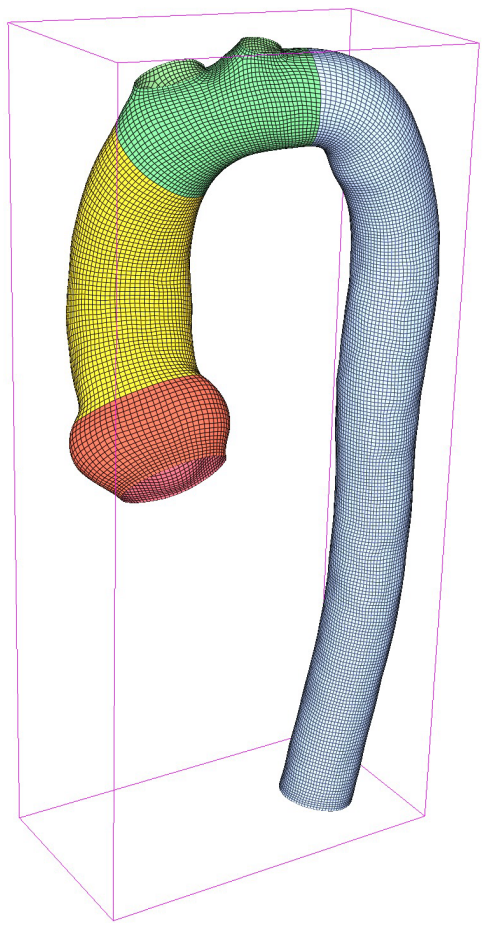

Figure 3. Deformed Quadrilateral Aorta Mesh
Aortic root (red); Ascending aorta (yellow); Aortic arch (green); Descending aorta (blue)

## 2.4 Simulation

The PyTorch-FEA library[24] was used for finite element (FE) simulation in this study to obtain the stress distribution on the aortic wall under the normal systolic blood pressure of 16kPa. 100 patients without aneurysms were selected from the INSPECT dataset, serving as the control group, and 37 patients with TAA were selected from the in-house dataset (Yale dataset). We implemented the Static Determinacy Approach (SDA) in the PyTorch-FEA library for FE simulations, which does not need patient-specific material properties. The details of SDA were reported in our previous work[15], where commercial software Abaqus was used. Because the FE meshes have node and element correspondence among patients, FE simulations are fully automatic with our open-source software PyTorch-FEA.

# 3. DISCUSSION

Table 1. nnU-Net Segmentation Performance

| | | DSC (%) | HD95 (mm) | Precision (%) | Recall (%) |
|---|---|---|---|---|---|
| nnU-Net | Average | 96.240 ±1.959 | 8.688 ±5.658 | 95.985 ±1.788 | 96.509 ±2.381 |
| | Max | 98.587 | 18.864 | 97.946 | 99.383 |
| | Min | 93.178 | 1.061 | 93.156 | 93.200 |

The nnU-Net model was developed using a composite dataset of 74 patients (47 from the Yale dataset and 27 from the HOCM dataset), randomly partitioned into 68 training samples and 6 independent testing samples. The network was trained for 1000 epochs using stochastic gradient descent with a Nesterov momentum of 0.99 and an initial learning rate of 0.01, which followed a "polyLR" decay schedule. All computational procedures were executed on NVIDIA H100 GPUs (80 GB memory). Performance was rigorously evaluated using the Dice Score Coefficient (DSC), Precision, Recall, and the 95th percentile Hausdorff Distance (HD95). While DSC, Precision, and Recall assess volumetric overlap (where higher values are optimal), the HD95 evaluates boundary adherence, with lower values denote precise boundary delineation. As shown in Table 1, the model's robust initial performance significantly alleviated the burden of subsequent manual refinement.

Table 2 Mesh Quality Evaluation

| | EQUIANGLE SKEW | ASPECT RATIO | SCALED JACOBIAN | MIN ANGLE | MAX ANGLE | SELF INTERSECTION |
|---|---|---|---|---|---|---|
| Mesh Quality | 0.0621 ± 0.0088 | 1.1726 ± 0.0633 | 0.9915 ± 0.0026 | 84.7219 ± 0.7557 | 95.3012 ± 0.763 | 0 |

Table 2 presents the mesh quality evaluation performed on the deformed quadrilateral meshes from a combined dataset of 147 patients (100 from the INSPECT dataset and 47 from the Yale dataset). The mesh quality performance was assessed using Equiangle Skew, Aspect Ratio, Scaled Jacobian, Min/Max Angles, and Self-Intersection checks. The Equiangle Skew, which quantifies deviation from an ideal equiangular shape (where 0 is optimal), demonstrated that the mesh elements were well-formed. The average Aspect Ratio was 1.1726, approximating the ideal value of 1.0 for quadrilateral elements. Similarly, The Scaled Jacobian quantifies the distortion in the transformation from an ideal reference element to the physical element, where values close to 1 indicate near-perfect, orthogonal geometry. The Minimum and Maximum Angle metrics revealed a tight distribution ranging from 84.72° to 95.30°, closely adhering to the ideal 90° standard. Crucially, no self-intersections were detected across the entire cohort. These high-quality mesh metrics are pivotal for ensuring numerical stability and accuracy in subsequent finite element simulations.

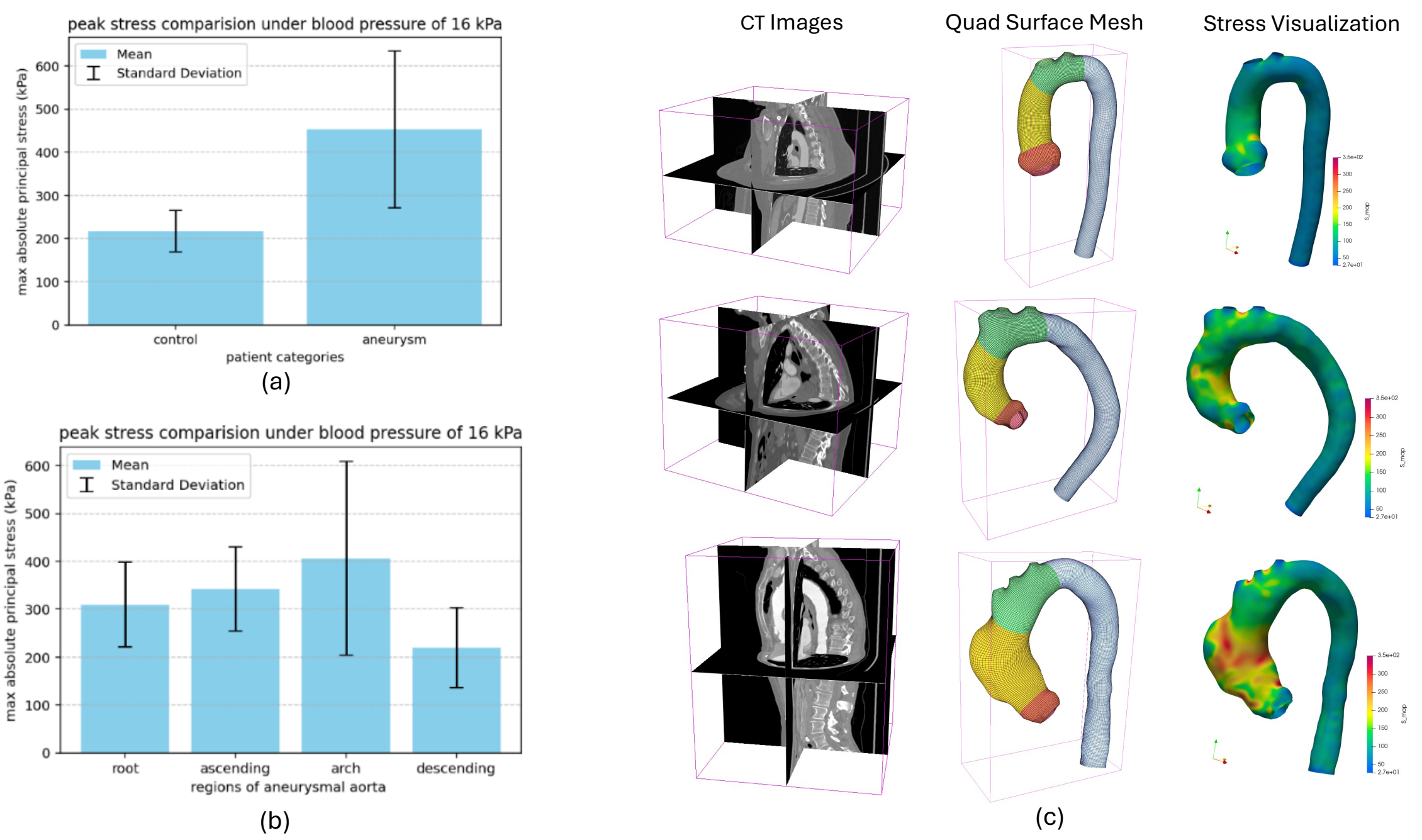

Figure 4. Comparison of stress distribution (left) and examples (right) from FE simulations

We compared stress analysis results between the non-aneurysmal patients (control group) and the 37 aneurysmal patients from the Yale dataset (aneurysm group). As shown in Figure 4a, aneurysmal patients exhibited both higher maximum absolute principal wall stress and greater standard deviation. These findings demonstrate that our pipeline can capture clinically relevant biomechanical differences that may aid in thoracic aortic aneurysm diagnosis.

We analyzed stress distribution across different aortic regions, including the aortic root, ascending aorta, aortic arch, and descending aorta. In the aneurysmal patients (Figure 4b), the distribution of maximum absolute principal stress showed the highest values in the aortic arch, followed by the ascending aorta, aortic root, and descending aorta. These findings suggest that our pipeline can reveal region-specific biomechanical insights, providing more detailed disease assessment and supporting personalized treatment planning. The findings are consistent with our previous works[15], which relied on a fully manual workflow.

Figure 4c shows pipeline outputs at the meshing and simulation stages for one non-aneurysm patient and two aneurysmal patients. The FE simulations show substantially increased wall stress concentrated within the dilated portions of the aorta in severe aneurysm cases, which implies stresses are risk indictor of TAA.

## 4. CONCLUSION

We present a comprehensive computational pipeline for patient-specific thoracic aortic aneurysm modeling, integrating segmentation, meshing, and finite element simulation. The framework highlighted clinically relevant biomechanical patterns, including elevated wall stress in aneurysmal patients and region-specific stress distributions. These findings emphasize the potential of our pipeline to enhance patient-specific disease assessment and support personalized treatment planning for thoracic aortic aneurysm.

**ACKNOWLEDGEMENT**: This work was supported in part by the NIH grant R01HL158829

# REFERENCES

[1] Shahoud, J. S., Miao, J. H. and Bolla, S. R., “Anatomy, Thorax, Heart Aorta,” [StatPearls], StatPearls Publishing, Treasure Island (FL) (2025).

[2] “WISQARS Leading Causes of Death Visualization Tool.”, Centers for Disease Control and Prevention, <https://wisqars.cdc.gov/lcd/> (18 August 2025 ).

[3] Kuzmik, G. A., Sang, A. X. and Elefteriades, J. A., “Natural history of thoracic aortic aneurysms,” Journal of Vascular Surgery 56(2), 565–571 (2012).

[4] Verstraeten, A., Luyckx, I. and Loeys, B., “Aetiology and management of hereditary aortopathy,” Nat Rev Cardiol 14(4), 197–208 (2017).

[5] Faggion Vinholo, T., Zafar, M. A., Ziganshin, B. A. and Elefteriades, J. A., “Nonsyndromic Thoracic Aortic Aneurysms and Dissections—Is Screening Possible?,” Seminars in Thoracic and Cardiovascular Surgery 31(4), 628–634 (2019).

[6] Nienaber, C. A., Clough, R. E., Sakalihasan, N., Suzuki, T., Gibbs, R., Mussa, F., Jenkins, M. P., Thompson, M. M., Evangelista, A., Yeh, J. S. M., Cheshire, N., Rosendahl, U. and Pepper, J., “Aortic dissection,” Nat Rev Dis Primers 2(1), 1–18 (2016).

[7] Wang, R., Lei, T., Cui, R., Zhang, B., Meng, H. and Nandi, A. K., “Medical image segmentation using deep learning: A survey,” IET Image Processing 16(5), 1243–1267 (2022).

[8] Conze, P.-H., Andrade-Miranda, G., Singh, V. K., Jaouen, V. and Visvikis, D., “Current and Emerging Trends in Medical Image Segmentation With Deep Learning,” IEEE Transactions on Radiation and Plasma Medical Sciences 7(6), 545–569 (2023).

[9] Chen, C., Qin, C., Qiu, H., Tarroni, G., Duan, J., Bai, W. and Rueckert, D., “Deep Learning for Cardiac Image Segmentation: A Review,” Front. Cardiovasc. Med. 7 (2020).

[10] Cao, L., Shi, R., Ge, Y., Xing, L., Zuo, P., Jia, Y., Liu, J., He, Y., Wang, X., Luan, S., Chai, X. and Guo, W., “Fully automatic segmentation of type B aortic dissection from CTA images enabled by deep learning,” European Journal of Radiology 121, 108713 (2019).

[11] Cheung, W. K., Bell, R., Nair, A., Menezes, L. J., Patel, R., Wan, S., Chou, K., Chen, J., Torii, R., Davies, R. H., Moon, J. C., Alexander, D. C. and Jacob, J., “A Computationally Efficient Approach to Segmentation of the Aorta and Coronary Arteries Using Deep Learning,” IEEE Access 9, 108873–108888 (2021).

[12] Saitta, S., Sturla, F., Caimi, A., Riva, A., Palumbo, M. C., Nano, G., Votta, E., Corte, A. D., Glauber, M., Chiappino, D., Marrocco-Trischitta, M. M. and Redaelli, A., “A Deep Learning-Based and Fully Automated Pipeline for Thoracic Aorta Geometric Analysis and Planning for Endovascular Repair from Computed Tomography,” J Digit Imaging 35(2), 226–239 (2022).

[13] Liang, X., Ebeida, M. S. and Zhang, Y., “Guaranteed-quality all-quadrilateral mesh generation with feature preservation,” Computer Methods in Applied Mechanics and Engineering 199(29–32), 2072–2083 (2010).

[14] Wisneski, A. D., Mookhoek, A., Chitsaz, S., Hope, M. D., Guccione, J. M., Ge, L. and Tseng, E. E., “Patient-Specific Finite Element Analysis of Ascending Thoracic Aortic Aneurysm,” J Heart Valve Dis 23(6), 765–772 (2014).

[15] Sun, C., Qin, T., Kalyanasundaram, A., Elefteriades, J., Sun, W. and Liang, L., “Biomechanical stress analysis of Type-A aortic dissection at pre-dissection, post-dissection, and post-repair states,” Computers in Biology and Medicine 184, 109310 (2025).

[16] Zheng, L., Chen, H., Qing, L., Zhuang, J., Meng, B. and Xu, X., “Automatic Segmentation of Aortic and Mitral Valves for Heart Surgical Planning of Hypertrophic Obstructive Cardiomyopathy,” Proceedings of the 15th Asian Conference on Machine Learning, 1715–1730, PMLR (2024).

[17] Huang, S.-C., Huo, Z., Steinberg, E., Chiang, C.-C., Lungren, M. P., Langlotz, C. P., Yeung, S., Shah, N. H. and Fries, J. A., “INSPECT: A Multimodal Dataset for Pulmonary Embolism Diagnosis and Prognosis,” arXiv:2311.10798 (2023).

[18] Isensee, F., Jaeger, P. F., Kohl, S. A. A., Petersen, J. and Maier-Hein, K. H., “nnU-Net: a self-configuring method for deep learning-based biomedical image segmentation,” Nat Methods 18(2), 203–211 (2021).

[19] Chen, J., Lu, Y., Yu, Q., Luo, X., Adeli, E., Wang, Y., Lu, L., Yuille, A. L. and Zhou, Y., “TransUNet: Transformers Make Strong Encoders for Medical Image Segmentation,” arXiv:2102.04306 (2021).

[20] Hatamizadeh, A., Nath, V., Tang, Y., Yang, D., Roth, H. R. and Xu, D., “Swin UNETR: Swin Transformers for Semantic Segmentation of Brain Tumors in MRI Images,” Brainlesion: Glioma, Multiple Sclerosis, Stroke

and Traumatic Brain Injuries, A. Crimi and S. Bakas, Eds., 272–284, Springer International Publishing, Cham (2022).

[21] Ma, J., He, Y., Li, F., Han, L., You, C. and Wang, B., "Segment anything in medical images," Nat Commun 15(1), 654 (2024).

[22] Chen, J., Qian, L., Wang, P., Sun, C., Qin, T., Kalyanasundaram, A., Zafar, M., Elefteriades, J., Sun, W. and Liang, L., "A 3D Image Segmentation Study of U-Net on CT Images of the Human Aorta with Morphologically Diverse Anatomy," 2024.10.02.616348 (2024).

[23] Fedorov, A., Beichel, R., Kalpathy-Cramer, J., Finet, J., Fillion-Robin, J.-C., Pujol, S., Bauer, C., Jennings, D., Fennessy, F., Sonka, M., Buatti, J., Aylward, S., Miller, J. V., Pieper, S. and Kikinis, R., "3D Slicer as an image computing platform for the Quantitative Imaging Network," Magnetic Resonance Imaging 30(9), 1323–1341 (2012).

[24] Liang, L., Liu, M., Elefteriades, J. and Sun, W., "PyTorch-FEA: Autograd-enabled finite element analysis methods with applications for biomechanical analysis of human aorta," Computer Methods and Programs in Biomedicine 238, 107616 (2023).